\documentclass[aps,prl,twocolumn]{revtex4}

\usepackage{graphicx}
\usepackage{amssymb}

\newcommand{\eps} {\varepsilon}
\newcommand{\nn} {\nonumber}

\begin{document}

\title{Discontinuous Meniscus Location in Tapered Capillaries Driven by Pressure
Difference and Dielectrophoretic Forces}

\author{Yoav Tsori\\
Department of Chemical Engineering, Ben-Gurion University of the
Negev, \\
P.O. Box 653, 84105 Beer-Sheva, Israel}

\date{\today}

\begin{abstract}

We calculate the meniscus location in tapered capillaries under the influence
of pressure difference and dielectrophoretic forces with and without gravity. We find
that the
meniscus location can be a discontinuous function of the pressure difference or
the applied voltage and that the meniscus can ``jump'' to one end or another of the
capillary. Phase diagrams are given as a function of the pressure and voltage,
depending on the geometrical parameters of the system. We further consider a
revision of the dielectric rise under dielectrophoretic force in wedge
capillaries and in the 
case of electrowetting, where the dielectrophoretic force is a
small perturbation. Finally, we also find discontinuous 
liquid--gas interface location in the case of liquid penetration into closed
volumes.
\end{abstract}

\maketitle


\section{1. Introduction} 

The location of the interface between two immiscible liquids or
between a liquid and a gas is essential 
in microfluidic
applications since the meniscus determines the way light is
scattered \cite{beebe,whitesides,quake}, how chemical species interact
\cite{manz,demello}, the way drops are transported in
small channels \cite{pgg_b_q,pgg_rmp}, and so forth. Attention must be given to liquid
channels or capillaries where the cross-section is nonuniform, since liquid
channels are never perfectly uniform and, as we show below, this nonuniformity
has a strong effect on the meniscus location \cite{tbj1,SB,tsori_langmuir2006}.

Consider a liquid contained in a channel with hard walls and a
varying cross section, and suppose that there is
a pressure difference $\Delta P$ between the two sides of the meniscus. At first
glance, it seems that an increase in $\Delta P$ will cause the
meniscus to move to the point where the Laplace pressure, as determined from the local geometry,
equals $\Delta P$,
so the meniscus location as a function of $\Delta P$ is continuous if the channel is smooth. When 
a dielectrophoretic force or, indeed, a gravitational force, is acting on the liquid at
the same time, we find that this naive picture is changed and that
 the
meniscus location becomes a discontinuous function of $\Delta P$.
The current case is different from those of refs \cite{SB,parry1} in that
(i) it takes into account gravity and electrostatic forces, (ii) pressure differences
exist even for a flat interface due to an external ``pumping'',
and (iii) the liquid layer is found underneath the solid surface and acts as a
big reservoir. 
\begin{figure}[h!]
\begin{center}
\includegraphics[scale=0.5,bb=115 190 530 785,clip]{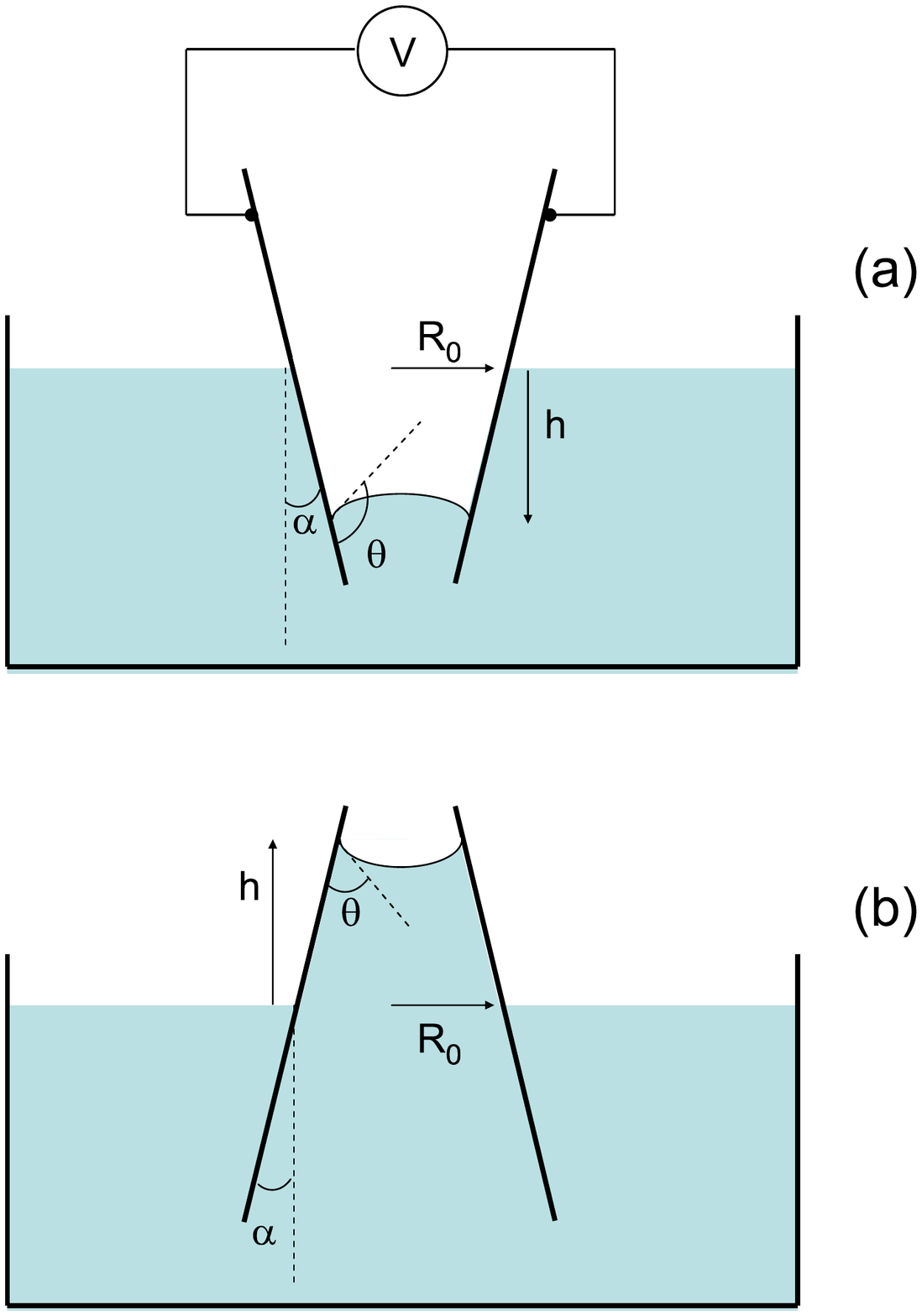}
\end{center}
\caption{\footnotesize{Schematic illustration of a tapered liquid channel made up of 
two tilted planes,
and definitions of parameters. Two of the possible cases are shown: (a) Hydrophobic
surface, $\cos\theta<0$, connected to a voltage supply. The opening angle
$\alpha$ is positive, and $h$ is negative. (b)
Hydrophilic surface, $\cos\theta>0$, negative $\alpha$ and positive $h$.
}}
\end{figure}

To be concrete, we consider the illustration in Figure 1. A similar wedge-shaped geometry
has recently been investigated on homogeneous (miscible) mixtures \cite{TTLnature, TLpre}.
Here, the meniscus
can be a liquid--gas or liquid--liquid interface, and its location 
is given by $h$ (being negative in part a and positive in part b).
The distance between the tilted planes at $h=0$ is $2R_0$, and $\alpha$ is the tilt angle.
In regular capillary
rise, $h$ is positive if the wetting angle $\theta$ is smaller than
$\frac12\pi$ and is negative if $\theta>\frac12\pi$. 
In addition to the gravitational forces and the pressure difference, a
dielectrophoretic force is acting: a high-frequency voltage difference $V$ is
imposed across the two
tilted planes. The frequency is assumed to be high enough so ionic screening does not
occur, and the force exerted on the liquid is dielectrophoretic in nature.

We restrict ourselves to narrow capillaries, where $\kappa
R\ll 1$ is satisfied, where $\kappa^{-1}=(\sigma/g\Delta\rho)^{1/2}$ is the
capillary length ($\sigma$ is the interfacial tension, and $\Delta\rho$ is the
density difference between the two liquids). In this case, as will be verified
below, the height
is larger than the radius, $h\gg R$, and the height variations of the
meniscus surface are negligible compared to the total height.
In mechanical equilibrium, at the contact line, the Laplace pressure 
is balanced by the hydrostatic pressure
\begin{eqnarray}\label{gov_eqn0}
P+\frac{\sigma}{r}=P_0-\Delta\rho gh+\frac12\Delta\eps E^2(h)
\end{eqnarray} 
where $P_0-P=\Delta P$ is the pressure difference and $r$ is the inverse curvature given
by
$r(h)=-R(h)/\cos(\theta+\alpha)$. Here, $2R(h)$ is the surface
separation at the meniscus' location and is given by $R(h)=R_0+h\tan\alpha$.
The height-dependent electric field is \cite{TLpre}
\begin{eqnarray}\label{Efield}
E(h)=\frac{V}{2\alpha}\frac{\sin\alpha}{R_0+h\tan\alpha} 
\end{eqnarray} 
Finally, $\Delta\eps$ is the permittivity difference between the liquid and the
gas or between the two liquids. 

The electric field has two general effects: (i) it exerts a net dielectrophoretic body
force on the liquid and (ii) it changes the contact angle. The exact influence of the
electric field thus depends on the applied frequency \cite{tbj2,tbj3}. Unless otherwise
stated, we will deal
with the high-frequency regime, where the electric field is
dielectrophoretic in nature, and the contact angle is unaffected by the field.
Equation \ref{gov_eqn0} is the basic relation for
the meniscus location, and it will be studied in detail for several cases
below.

\section{2. No Gravitational Force} 

The gravitational force is zero if the channel is horizontal, as occurs in many
cases, or if $\Delta\rho$ is sufficiently small. 
In this case, we are faced with the following equation:
\begin{eqnarray}\label{gov_eqn2}
P+\frac{\sigma}{r}=P_0+\frac12\Delta\eps E^2(h)
\end{eqnarray} 
Note that if $\Delta P=0$ is zero, there is a
balance of the dielectrophoretic force against the surface tension, and this
leads to some liquid height $h$ where the forces balance. $h$ in this
case is a continuous function of the system parameters (e.g., $V$). Since
this is the less interesting scenario, we now assume, 
without loss of generality, that $\Delta P=P_0-P>0$. We obtain the
following for the mechanical balance:
\begin{figure}[h!]
\begin{center}
\includegraphics[scale=0.55,bb=100 180 490 615,clip]{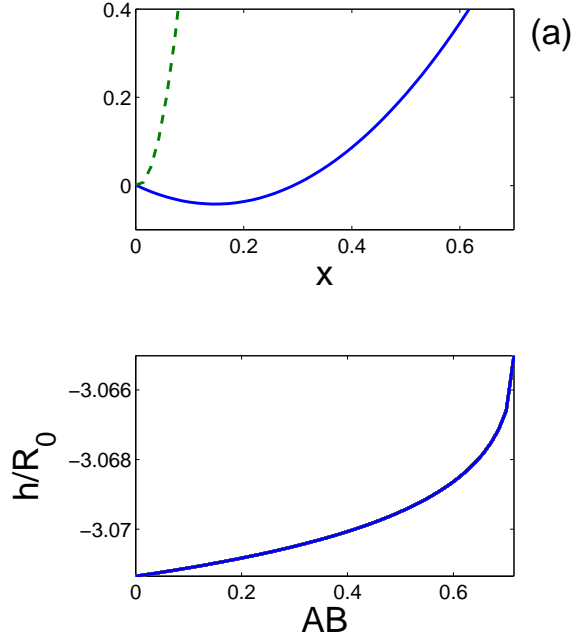}
\end{center}
\caption{\footnotesize{(a) Plot of the left-hand side of eq \ref{parabola_x} versus $x$.
If
$A$ is large (dashed green curve), there is no real root to the equation. For smaller
values of $A$ (solid blue curve), the parabola first descends before it ascends and there
are two roots. (b) Plot of the meniscus height $h$ normalized by $R_0$ as a
function of $AB$. At larger values of $AB$,
$AB>(AB)_c$, the meniscus jumps to
the top of the
capillary. In both parts, $\alpha=0.1\pi$, $\theta=0.6\pi$, $R_0=1$
mm and $(AB)_c=0.71$.}
}
\end{figure}
\begin{eqnarray}
\frac18\Delta\eps
V^2\left(\frac{\sin\alpha}{\alpha}
\right)^2+\sigma\cos(\theta+\alpha)\left(R_0+h\tan\alpha\right)+\nn\\
\Delta P\left(R_0+h\tan\alpha\right)^2=0~~~~ 
\end{eqnarray} 
This relation can be further cast in the dimensionless form:
\begin{eqnarray}\label{parabola_x}
\frac18 B\left(\frac{\sin\alpha}{\alpha}
\right)^2+\cos(\theta+\alpha)x+Ax^2=0 
\end{eqnarray} 
where 
\begin{eqnarray}
x\equiv1+\frac{h}{R_0}\tan\alpha
\end{eqnarray} 
and the dimensionless numbers are
\begin{eqnarray}
A&=&\frac{\Delta PR_0}{\sigma} \nonumber\\
B&=&\frac{\Delta\eps V^2}{\sigma R_0}
\end{eqnarray} 
Let us look at the magnitude of $A$ and $B$. For a pressure difference $\Delta
P=1$ atm, surface tension $\sigma=0.1$ N/m, and $R_0=1$ mm, we find $A=10^3$.
If $\Delta\eps=10\eps_0$ (where $\eps_0$ is the vacuum permittivity) and
$V=100$V, we find $B=10^{-2}$. Thus, $A$ can be very large while $B$ is
typically quite small, and the product is $AB\gtrsim 1$.

Equation \ref{parabola_x} is a parabola in the variable $x$, and 
$x$ is always positive, because $R_0+h\tan\alpha>0$. Let us look at the case of
positive
$A$ and $B$ but negative $\cos(\theta+\alpha)$; this means that the dielectric
liquid is pulled toward higher values of $h$ by the electric field and applied
pressure. The minimum of the parabola is at $x_0=-\cos(\theta+\alpha)/2A$.
\begin{figure}[h!]
\begin{center}
\includegraphics[scale=0.4,bb=35 175 540 580,clip]{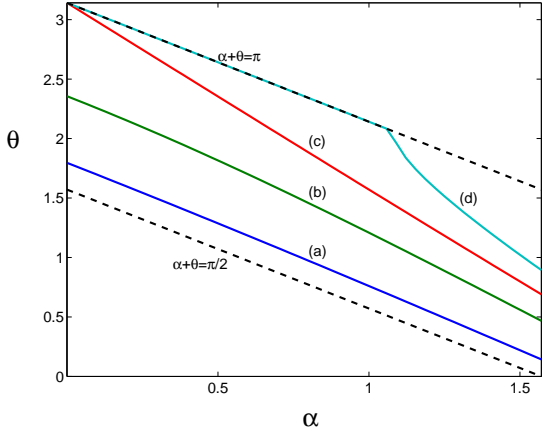}
\end{center}
\caption{\footnotesize{Phase diagram in the $\alpha$ -- $\theta$ plane. $\alpha$ and
$\theta$
are restricted to be between the two diagonal dashed lines $\alpha+\theta=\pi$
and $\alpha+\theta=\frac12\pi$. The solid line marked as (a), corresponding
 to $AB=0.1$, divides the plane into two parts: below it, the meniscus is at the
top of the capillary, and above it, $h$ is a continuous function of $\alpha$
and $\theta$. Lines (b), (c), and (d) are the same, but with $AB=1$, $AB=2$,
and $AB=3<(AB)^*$, respectively. At $AB>(AB)^*=\frac12\pi^2$, the meniscus is
at the top, irrespective of $\alpha$ and $\theta$.
}}
\end{figure}
If $B$ or $A$ is sufficiently small, there are
two roots to the equation: an unstable one $x_1>x_0$ and a stable root
$x_2<x_0$ (see Figure 2a). Figure 2b shows $h$ as a function of the
product $AB$ at given $\alpha$ and $\theta$ values.
There is no solution if $A$ or $B$ is too large, and this occurs when $AB>(AB)_c$, where
\begin{eqnarray}\label{AB_crit}
(AB)_c=2\cos^2(\alpha+\theta)/(\sin\alpha/\alpha)^2
\end{eqnarray}
is a critical value of $AB$. Hence, an increase of $AB$ from below to above $(AB)_c$
leads to a jump of the meniscus from $h_0$ given by
\begin{eqnarray}
h_0=-\frac{R_0}{\tan\alpha}\left(1+\frac{1}{2A}\cos(\theta+\alpha)\right) 
\end{eqnarray} 
to the top of the capillary. Relation \ref{AB_crit} can be inverted to give a condition
for a critical angle $\theta_c$:
\begin{eqnarray}
\theta_c=\arccos\left(\sqrt{\frac12 AB}\frac{\sin\alpha}{\alpha}\right)-\alpha
\end{eqnarray}
The meniscus location is changing continuously for $\theta>\theta_c$ and is found at the capillary's top when $\theta<\theta_c$.

The phase diagram is presented in
Figure 3 as a function of $\alpha$ and $\theta$ for different values of $AB$.
As the value of $AB$ increases from zero, the area below the $\theta_c(\alpha)$ lines (solid curves)
increases. This area is where the meniscus is found at the top of the
liquid channel, while above it the meniscus location is continuous.
Lines (a) and (b) are for $AB=0.1$ and $AB=1$, respectively.
Curve (c) corresponds to $AB=2$, and it is the first that touches the limiting diagonal
line $\alpha+\theta=\pi$. 
As $AB$ increases above $2$, the line $\theta_c(\alpha)$ overlaps with the
$\alpha+\theta=\pi$ line for small values of $\theta$ but not for large ones
(curve (d) for $AB=3$). The maximum value of
$2\cos^2(\alpha+\theta)/(\sin\alpha/\alpha)^2$ occurs at $\alpha=\theta=\frac12\pi$.
Therefore, from eq \ref{AB_crit}, we see that
there is another special value of $AB$: 
\begin{eqnarray}
(AB)^*=\frac12\pi^2
\end{eqnarray}
Hence, when $AB>(AB)^*$ the meniscus is at the top for all values of
$\alpha$ and $\theta$.

In the following section, we investigate the case of a tapered
liquid channel under a gravitational force in addition to the
dielectrophoretic force but in the absence of pressure difference.

\section{3. Gravity Effect in the Absence of Pressure Difference}

We now turn to the rather complex case where there are dielectrophoretic and
interfacial tension forces acting, together with gravity, but no pressure
difference. 
The governing equation reads as follows:
\begin{eqnarray}\label{gov_eqn3}
\frac18\Delta\eps V^2\left(\frac{\sin\alpha}{\alpha}\right)^2+
\sigma\cos(\alpha+\theta)\left(R_0+h\tan\alpha\right)=\nn\\
\Delta\rho gh\left(R_0+h\tan\alpha\right)^2~~~~
\end{eqnarray}
The high-frequency limit of the potential $V$ corresponds to the purely
dielectrophoretic case, where $\theta$ is simply the zero-field contact angle.
In the low-frequency limit, $\omega\to 0$, the dielectrophoretic force vanishes
(this is equivalent to setting $V=0$ above), while $\theta=\theta(V)$ is the
voltage-dependent (Lipmann) contact angle.

The gravitational force introduces the capillary length
$\kappa^{-1}=(\sigma/g\Delta\rho)^{1/2}$, and this length is used to scale all lengths in
the system. The above equation can be expressed using dimensionless lengths as follows:
\begin{eqnarray}\label{gov_eqn3b}
\cos(\alpha+\theta)&=&f(\bar{h})\\
f(\bar{h})&\equiv&\bar{R}_0\bar{h}
+\tan\alpha\bar{h}^2-\frac18\frac{B\left(\frac{\sin\alpha}{\alpha}\right)^2}{1+
\frac{\bar{h}}{\bar{R}_0}\tan\alpha}\nn
\end{eqnarray}
where 
\begin{eqnarray}
\bar{h}=\kappa h,~~~~~\bar{R}_0=\kappa\bar{R}_0
\end{eqnarray}
To investigate the field effect,  we note
that $B$ is usually small and
seek solutions with $B\ll 1$. In the absence of an electric 
field ($B=0$), eq \ref{gov_eqn3b} was recently studied in ref.
\cite{tsori_langmuir2006}, and we give here a brief summary of the main results.
We rewrite eq \ref{gov_eqn3b} without the $B$-dependent terms as follows:
\begin{eqnarray}
\cos(\alpha+\theta)&=&f(\bar{h})\nonumber\\
f(\bar{h})&=&\bar{R}_0\bar{h}
+\tan\alpha\bar{h}^{2}\label{gov_eqn3c}
\end{eqnarray}
$f(\bar{h})$ has a maximum at $\bar{h}=\bar{h}^*=-\bar{R}_0/(2\tan\alpha)$, and
its value at the maximum is $f(\bar{h}^*)=-\bar{R}_0^2/(4\tan\alpha)$. Thus,
for hydrophilic surfaces ($\cos(\alpha+\theta)>0$), for small and negative
$\alpha$ values, there is a solution to eq \ref{gov_eqn3c}. If $|\alpha|$ is too
large, however, there is no solution: $-\bar{R}_0^2/(4\tan\alpha)$ is smaller
than $\cos(\alpha+\theta)$, and the meniscus jumps from 
\begin{eqnarray}
\bar{h}^*=-\frac{\bar{R}_0}{2\tan\alpha}
\end{eqnarray}
to the capillary's end (to the top if $\alpha<0$ and to the bottom if
$\alpha>0$).
The condition for the meniscus' jump is as follows:
\begin{figure}[h!]
\begin{center}
\includegraphics[scale=0.65,bb=145 180 450 620,clip]{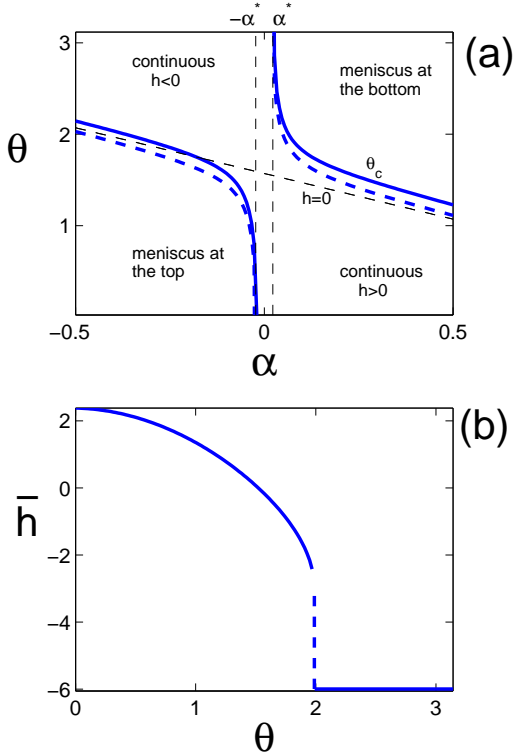}
\end{center}
\caption{\footnotesize{(a) Phase diagram in the $\alpha$ -- $\theta$ plane a of a liquid
in a tapered
channel with dielectrophoretic and gravitational forces. The thick dashed lines are
$\theta_c$ from eq \ref{crit_th_field} with $B=0$, and the thick solid lines are for
the case where $B=0.5$. Above $\theta_c$ and for positive $\alpha$, the
meniscus is at the capillary's bottom, whereas it is continuously varying below
$\theta_c$. For negative $\alpha$, the meniscus is at the top provided that
$\theta<\theta_c$. The thin diagonal dashed line represents the meniscus
height $h=0$; above it, the meniscus is at a negative location, and below it, the
meniscus is at a positive location. (b) Normalized meniscus height $\bar{h}$ as
a function of $\theta$ for a fixed  value $\alpha=0.05$. When $\theta$ becomes
larger than $\theta_c$ (in this case, $\theta_c\simeq 2$), the meniscus jumps to
the capillary's bottom. In (a) and (b), we took $\bar{R}_0=0.3$.}
}
\end{figure}
\begin{eqnarray}\label{cond_zero_field}
\cos\left(\alpha+\theta\right)=f(\bar{h}^*)=-\frac14\frac{\bar{R}_0^2}{
\tan\alpha}~~~~~~~~~{\rm no~field}
\end{eqnarray} 
Alternatively, the condition can be expressed as a condition for a critical
angle $\theta_c(\alpha)$:
\begin{eqnarray}
\theta_c=\arccos\left(-\bar{R}_0^2/4\tan\alpha\right)-\alpha~~~~~~~~~{\rm
no~field}
\end{eqnarray} 
For a negative angle $\alpha$, the meniscus location is continuous if
$\theta>\theta_c$, while the meniscus is at the top for every
$\theta<\theta_c$. 
The system is invariant with respect to the transformation
$\alpha\to -\alpha$ and $\theta\to\pi-\theta$. Hence, for a positive value of
$\alpha$, if $\theta>\theta_c$, the meniscus is at the capillary's bottom, and
its location is continuously changing for $\theta<\theta_c$. There is a
``special'' angle $\alpha^*$ given by $\alpha^*=\arcsin(\bar{R}_0^2/4)$. If
$-\alpha^*<\alpha<\alpha^*$, the meniscus location as a function of $\theta$ is
continuous for all $\theta$ values.

We now add the field's effect and treat eq \ref{gov_eqn3b} perturbatively
with small $B$ values. We are looking for the maximum of
$f(\bar{h})$ from eq \ref{gov_eqn3b}. Using $B\ll 1$, we find that
\begin{eqnarray}
\bar{h}^*=-\frac{\bar{R}_0}{2\tan\alpha}-\frac14\frac{
B\left(\sin\alpha/\alpha\right)^2/\bar{R}_0}{
1+B\left(\sin\alpha/\alpha\right)^2\tan\alpha/\bar{R}_0^2}
\end{eqnarray} 
To first order in $B$, the relation replacing eq \ref{cond_zero_field} is 
\begin{eqnarray}
\cos\left(\alpha+\theta\right)=-\frac14\frac{\bar{R}_0^2}{
\tan\alpha}-\frac14 B\left(\frac{\sin\alpha}{\alpha}\right)^2
\end{eqnarray} 
Note that, in the preceding derivation, the second term on the right-hand side
was assumed to be small compared with
the first term. However, as $\alpha$ approaches $\frac12\pi$, the first term
goes to zero, while $\sin\alpha/\alpha$ is finite: in this limit, the
dielectrophoretic term dominates and does not represent a small correction anymore, and
our derivation fails.

The condition for the critical angle becomes
\begin{eqnarray}\label{crit_th_field}
\theta_c=\arccos\left(-\frac14\frac{\bar{R}_0^2}{\tan\alpha}-\frac14
B\left(\frac{\sin\alpha} {\alpha}\right)^2\right)-\alpha
\end{eqnarray} 
Figure 4 summarizes these findings. The zero-field critical value of
$\theta(\alpha)$ is shown as dashed lines in Figure 4a, while the solid lines
represent the critical angle for the case where $B=0.3$. At a positive $\alpha$ value,
if $\theta<\theta_c$, the meniscus location is continuously changing with
$\theta$. Below the diagonal dashed line $h$, is positive, while $h<0$ above the dashed
line.
Above the $\theta_c$, line the meniscus is at the capillary's bottom. Note that
the electric field breaks the symmetry of the system: while basically similar
behavior appears at negative values of $\alpha$, the operation $\alpha\to
-\alpha$ and $\theta\to\pi-\theta$ does not leave the system invariant as it
does for the $B=0$ case.

\section{4. Revision of the Problem of Dielectric Rise}

\begin{figure}[h!]
\begin{center}
\includegraphics[scale=0.35,bb=25 180 545 585,clip]{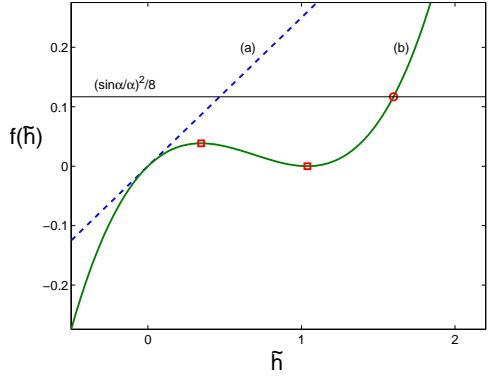}
\end{center}
\caption{\footnotesize{Graphical solution of eq \ref{gov_eqn1}. The diagonal dashed
line
(a) represents $f(\tilde{h})$ for the case where the opening angle is $\alpha=0$
(regular capillary rise). The curve (b) is $f(\tilde{h})$ for the case where
$\alpha=-\pi/7$. The
horizontal line is $(\sin\alpha/\alpha)^2/8$. Their
intersection occurs at the marked circle. As a function of $\alpha$, the
solution may
jump from the right branch of $f(\tilde{h})$ to the left one. The two
squares mark the extrema $\tilde{h}_1$ and $\tilde{h}_2$ from eq
\ref{extrema_h1h2} (see text).}
}
\end{figure}
\begin{figure}[h!]
\begin{center}
\includegraphics[scale=0.55,bb=120 175 485 615,clip]{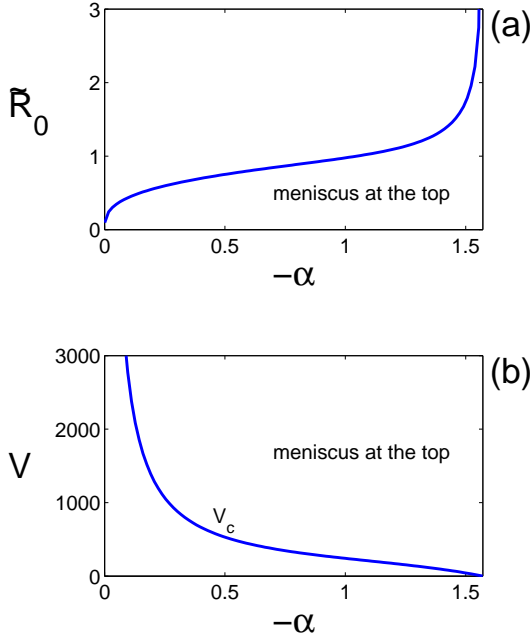}
\end{center}
\caption{\footnotesize{(a) Phase diagram for the liquid rise in tapered capillaries with
zero
pressure difference ($\Delta P=0$) and negligible interfacial tension. 
The opening
angle $\alpha$ is assumed to be negative here. At a given value of $\alpha$, if
$\tilde{R}_0$ is smaller than its value given by eq \ref{gov_diag1} (solid
curve), the meniscus is found at the top of the capillary. (b)
Phase diagram in the $\alpha$-$V$ plane. As $V$ (in volts) is increased above
the $V_c$ value given by eq \ref{Vc} (solid curve), the meniscus jumps to the
capillary's top. We took Earth's gravity, the density of water, $R_0=0.1$ mm, and
$\Delta\eps$ equals five times the vacuum permittivity.}
}
\end{figure}

We now consider the case of a liquid dielectric rise in tapered capacitors. We
assume that the interfacial tension $\sigma$ is negligibly
small and that there are no imposed pressure differences: $\Delta P=P_0-P=0$. 
The first treatment to this problem was given by T. B. Jones in a classical paper
from 1974 \cite{tbj1}.
The hydrostatic pressure is balanced by the dielectrophoretic force, yielding
\begin{eqnarray}
\frac18\Delta\eps V^2\left(\frac{\sin\alpha}{\alpha}\right)^2=\Delta\rho
gh\left(R_0+h\tan\alpha\right)^2 
\end{eqnarray} 
Or, expressed differently,
\begin{eqnarray}
\frac18\left(\frac{\sin\alpha}{\alpha}\right)^2&=&f(\tilde{h})\nonumber\\
f(\tilde{h})&=&\tilde{h}\left(\tilde{R}_0+\tilde{h}\tan\alpha\right)^2
\label{gov_eqn1}
\end{eqnarray} 
where the dimensionless quantities are
\begin{eqnarray}
\tilde{h}=h/L,~~~~~\tilde{R}_0=R_0/L
\end{eqnarray} 
and $L$ is given by
\begin{eqnarray}
L=\left(\frac{\Delta\eps V^2}{\Delta\rho g}\right)^{1/3} 
\end{eqnarray} 
The difference between our derivation and the one by Jones is the expression
for the electric field (eq \ref{Efield}), which is more accurate for large
angles ($\alpha$); however, in the limit $\alpha\to 0$, we expect to recover Jones'
results. 
In the more usual case of a nontapered capillary ($\alpha=0$), we recover
the familiar expression
\begin{eqnarray}
\tilde{h}=\frac{1}{8\tilde{R}_0^2} 
\end{eqnarray} 
which is equivalent to 
\begin{eqnarray}
h=\frac{\Delta\eps V^2}{8g\Delta \rho}\frac{1}{R_0^2} 
\end{eqnarray} 
in physical quantities.

Let us concentrate on the case where $\alpha<0$. As is seen in Figure 5,
$f(\tilde{h})$ has an inflection point. There are two extrema located at
\begin{eqnarray}
\tilde{h}_2&=&-\frac13\frac{\tilde{R}_0}{\tan\alpha}\nonumber\\
\tilde{h}_1&=&-\frac{\tilde{R}_0}{\tan\alpha} \label{extrema_h1h2}
\end{eqnarray} 
Both extrema are positive and smaller than the maximum height in the capillary,
$\tilde{h}_{\rm max}=-\tilde{R}_0/\tan\alpha$. The value of $f(\tilde{h})$ at these
extrema is
\begin{eqnarray}
f(\tilde{h}_1)&=&0\nonumber\\ 
f(\tilde{h}_2)&=&-\frac{4}{27}\frac{\tilde{R}_0^3}{\tan\alpha} 
\end{eqnarray} 
The left-hand side of eq
\ref{gov_eqn1} represents a horizontal line, and it can cross
$f(\tilde{h})$ at the right or left branch. On increasing $\alpha$ from zero,
the solution jumps from the right branch to the left one when
$\frac18\left(\frac{\sin\alpha}{\alpha}\right)^2=f(\tilde{h}_2)$, that is, when 
\begin{eqnarray}\label{gov_diag1}
\left(\frac{\sin\alpha}{\alpha}\right)^2=-\frac{32}{27}\frac{\tilde{R}_0^3}{
\tan\alpha}
\end{eqnarray} 
Alternatively, for the critical voltage $V_c$, we find the following expression:
\begin{eqnarray}\label{Vc}
V_c=-\frac{32}{27}\frac{g\Delta\rho}
{\Delta\eps}\frac{R_0^3}{\tan\alpha\left(\frac{\sin\alpha}{\alpha}\right)^2}
\end{eqnarray} 

Figure 6a shows the phase diagram for the meniscus location. The solid line is
$\tilde{R}_0$ from eq \ref{gov_diag1}. At a given negative value of $\alpha$, if
$\tilde{R}_0$ is below the critical line, the meniscus is at the top of the
capillary, while above this line the meniscus location is continuous. Figure 6b
expresses this behavior in the $\alpha$ -- $V$ plane. As $V$ is increased above $V_c$
given
by eq \ref{Vc}, the meniscus jumps to the capillary's top. The meniscus
location is continuous below it. 

In the following section, we find that similarities appear for a liquid channel
blocked at one end, because in this case the pressure difference $\Delta P$ depends on
the meniscus location.


\section{5. Liquid Penetration into Closed Volumes}

In this section, we turn to describing the meniscus location in tapered 
channels where one
of the ends is blocked. The interface is a gas--liquid interface, as 
depicted in Figure 7. 
In the following, we assume that gravity and electric fields are 
absent, but nonetheless we find
that the meniscus location can be discontinuous. The reason for this is 
the nonlinear dependence of the
gas pressure in the area enclosed by the liquid and the walls 
on the
gas volume, and the gas pressure must be balanced by the Laplace pressure. 
\begin{figure}[h!]
\begin{center}
\includegraphics[scale=0.85,bb=185 540 380 745,clip]{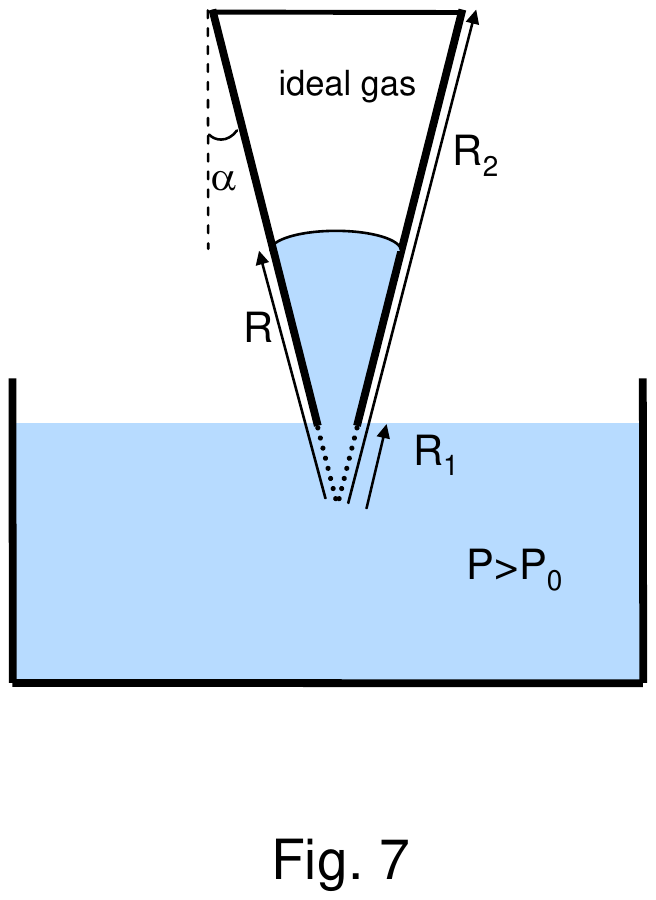}
\end{center}
\caption{\footnotesize{Schematic illustration of liquid penetration into closed volumes.
The
wedge-shaped channel is closed at one end and is brought into contact with the 
liquid
reservoir. Before contact, the pressure is equal to $P_0$ everywhere. After contact, 
the liquid
may penetrate into the channel if the outside pressure is increased to 
$P=P_0+\Delta P$. 
The larger and smaller channel 
radii are $R_2$ and $R_1$, respectively, measured from the imaginary meeting 
point of the walls, and the penetration depth is $R$.}}
\end{figure}

The liquid channel is wedge-shaped, with large and small opening radii $R_2$ and $R_1$, as
measured from the imaginary meeting point of the channel walls, and the 
opening angle is $2\alpha$, as defined in Figure 7.
Let us call $V_0$ and $P_0$ the gas volume and pressure, respectively, just before contact
of the liquid channel with the liquid reservoir; $P_0$ is also the
ambient pressure in the liquid.
The channel is then brought in contact with the liquid,
the liquid pressure is increased by an amount $\Delta P$,  and the liquid may penetrate to
a distance $R$
inside the capillary. It follows from simple geometry that the maximum gas volume
$V_0$ (i.e., when $R=R_1$) is 
\begin{eqnarray}
V_0=L_z\alpha\left(R_2^2-R_1^2\right)
\end{eqnarray} 
where $L_z$ is its depth in the third dimension (in the page).
The total gas volume for $R>R_1$ is
\begin{eqnarray}
V=L_z\alpha\left(R_2^2-R^2\right)
\end{eqnarray} 
\begin{figure}[h!]
\begin{center}
\includegraphics[scale=0.45,bb=90 175 510 590,clip]{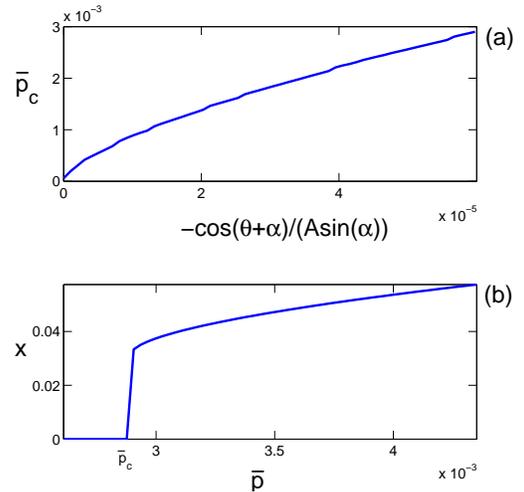}
\end{center}
\caption{\footnotesize{(a) Plot of $\bar{p}_c$, the critical value of the reduced
pressure 
$\bar{p}=\Delta
P/P_0$, as a function of $\cos(\alpha+\theta)/(A\sin\alpha)$, where 
$A=R_2P_0/\sigma$. $\bar{p}_c$ is computed from eq 
\ref{eps_c}.
(b) Scaled meniscus location $x=R/R_2$ as a function of the scaled 
pressure difference $\bar{p}$. If $\bar{p}<\bar{p}_c$, the meniscus 
is at $x=0$; that is, the liquid does not penetrate into the channel. 
When $\bar{p}=\bar{p}_c$, the meniscus jumps to $x=x_0$ as given by 
eq \ref{men_loc_jump} with 
$\cos(\alpha+\theta)/(A\sin\alpha)=-6\times 10^{-5}$.}
}
\end{figure}
Assuming the ideal gas law $P_0V_0=PV$, we find that 
the gas pressure is
\begin{eqnarray}
P=P_0\frac{R_2^2-R_1^2}{R_2^2-R^2}
\end{eqnarray} 
The outside pressure is increased to $P_0+\Delta P$. We continue in the limit of the
very small channel opening, that is, $R_1\ll R_2(\Delta P/P_0)^{1/2}$ and $R_1\ll R$, and
therefore, eq \ref{gov_eqn2} with $E=0$ leads to 
\begin{eqnarray}
P_0\frac{R^2}{R_2^2-R^2}=\frac{\sigma\cos(\alpha+\theta)}{R\sin\alpha}+\Delta P
\end{eqnarray} 
This governing equation can be written using the dimensionless variables
\begin{eqnarray}
x&=&\frac{R}{R_2}\nonumber\\
A&=&\frac{R_2P_0}{\sigma}\nonumber\\
\bar{p}&=&\frac{\Delta P}{P_0}
\end{eqnarray}
in the following form:
\begin{eqnarray}\label{gov_eqn4}
\frac{1}{A}\frac{\cos(\alpha+\theta)}{\sin\alpha}&=&f(x)\\
f(x)&=&-\bar{p} x+\frac{1}{A}\frac{\cos(\alpha+\theta)}{\sin\alpha}x^2+(1+\bar{p})x^3\nn
\end{eqnarray}
At this point, we
would like to remind the reader that if the ambient pressure is
atmospheric, $P_0=10^5$ Pa, it follows
that $A=P_0R_2/\sigma\approx 10^4 \gg 1$, where we took $\sigma=0.1$ 
N/m and
$R_2=1$ cm. In the following, we will concentrate on $\bar{p}\ll 1$; 
however, we will assume $A^{-1}\ll\bar{p}$. 

We seek the meniscus location $x=R/R_2$ for a given pressure difference $\Delta P$ and for
hydrophobic channels. This means that we seek the solution of eq \ref{gov_eqn4} for fixed
$A$, negative $\cos(\alpha+\theta)$, and increasing $\bar{p}$. $x$ is nonnegative and
smaller than $1$, and the function $f(x)$ descends for small $x$ values and ascends for
larger $x$ values. The minimum of $f(x)$ is obtained at $x_0$ given by
\begin{eqnarray}
x_0=\frac{-\frac{\cos(\alpha+\theta)}{\sin\alpha}+\sqrt{\frac{
\cos^2(\alpha+\theta)}{\sin^2\alpha}
+3A^2\bar{p}(1+\bar{p})}}{3A(1+\bar{p})}
\end{eqnarray}
and the value at the minimum is $f(x_0)$. When $f(x_0)>\cos(\alpha+\theta)/(A\sin\alpha)$,
the meniscus is at $x=0$; that is the liquid does not penetrate into the closed volume.
If $f(x_0)<\cos(\alpha+\theta)/(A\sin\alpha)$, then the meniscus is at $x>x_0$, as is
given by a direct
solution of eq \ref{gov_eqn4}. The meniscus jumps from $x=0$ to $x=x_0$
when $f(x_0)=\cos(\alpha+\theta)/(A\sin\alpha)$. This can be expressed by the rather long
expression 
\begin{eqnarray}
\frac{\cos(\alpha+\theta)}{A\sin\alpha}=\left[2\left(\frac{\cos(\alpha+\theta)}{
A\sin\alpha}\right)^3-\right.\nn\\
2\left(\frac{\cos(\alpha+\theta)}{A\sin\alpha}\right)^2\sqrt{
\left(\frac{\cos(\alpha+\theta)}{A\sin\alpha}\right)^2
+3\bar{p}(1+\bar{p})}-\nn\\
6\bar{p}(1+\bar{p})\sqrt{\left(\frac{\cos(\alpha+\theta)}{A\sin\alpha}\right)^2
+3\bar{p}(1+\bar{p})}+\nn\\
\left. 9\bar{p}(1+\bar{p})\frac{\cos(\alpha+\theta)}{A\sin\alpha}\right]/
\left(27(1+\bar{p})^2\right)~~~~
\end{eqnarray}

As we mentioned above, we are interested in the limit $A^{-1}\ll\bar{p}\ll 1$, and in this
case, we can obtain the following approximate and much simpler
expression for the critical value of the dimensionless pressure $\bar{p}_c$:
\begin{eqnarray}\label{eps_c}
-\frac{2}{\sqrt{3}}\frac{\bar{p}_c\sqrt{\bar{p}_c(1+\bar{p}_c)}}{3+2\bar{p}_c}=\frac1A\frac{\cos(\alpha+\theta)}{\sin\alpha}
\end{eqnarray}
As $\bar{p}$ is increased from zero to $\bar{p}=\bar{p}_c$, the meniscus jumps from $x=0$ to $x=x_0$, where $x_0$ is given by:
\begin{eqnarray}\label{men_loc_jump}
x_0\approx\sqrt{\frac{\bar{p}_c}{3(1+\bar{p}_c)}}-
\frac1A
\frac{\cos(\alpha+\theta)/\sin\alpha}{3(1+\bar{p}_c)}
\end{eqnarray}

Figure 8a shows $\bar{p}_c$ as a function of increasing $A^{-1}$ as given by
eq \ref{eps_c}. $\bar{p}_c\to 0$ if $A^{-1}\to 0$, and increases monotonically with
$A^{-1}$. Figure 8b shows the scaled meniscus location $x$ as a function of $\bar{p}$
at a given value of $A$. As $\bar{p}$ is increased from zero, $x$ jumps from $x=0$ to
$x=x_0$ at $\bar{p}=\bar{p}_c$ and increases monotonically with a further increase in
$\bar{p}$.

We would like to stress that one can also ask the following question:
at what value of 
interfacial tension $\sigma$ does the meniscus jump for a given value 
of $\bar{p}$? The expression for $A_c$ as a function of $\bar{p}$ is
obtained directly from the inversion of eq \ref{eps_c}.
In addition, we recall that the ideal gas pressure $P_0$ 
is given by $P_0=nk_BT$, where $n$ is the gas density, $k_B$ is the 
Boltzmann constant, and $T$ is the absolute temperature.
We thus point out that one can hold both $\Delta P$ and $\sigma$ 
constant while changing the temperature. Again, eq \ref{eps_c} can 
be used to find a critical temperature $T_c$ for the meniscus jump.

\section{6. Conclusions}

In this article, we consider in detail the location of a liquid--liquid or 
liquid--gas interface in tapered capillaries. The driving forces are
external pressure difference, interfacial tension, and electrostatic 
and gravitational forces. 

As one would naively expect, a small change in the external forces
usually leads to a small change in the equilibrium 
interface location. However, as is shown above without exception, 
due to the nonlinearity of the competing forces, there are critical 
values of the external parameters: pressure, wetting 
angle, voltage, and so forth. If the 
external force is close to its critical value, the equilibrium 
interface location is discontinuous. This is a 
rather general phenomenon in capillaries with nonuniform cross-sections, and occurs even
at small ``opening angles'' $\alpha$.
In section 2, we considered a capillary with a pressure difference and
dielectrophoretic forces and obtained the threshold values of applied 
voltage or pressure difference to drive the meniscus location 
discontinuity. In section 3, we studied the liquid rise in capillaries
under a weak dielectrophoretic force and gravity. The meniscus jump was 
discussed in terms of the wetting angle $\theta$ and wedge opening 
angle $\alpha$. We further looked in section 4 at the classical 
problem 
of dielectric rise, but this time in a tapered capacitor. Again, we 
found that the meniscus location exhibits discontinuities as a function 
of applied voltage or the geometrical parameters. Finally, in section 5, 
we considered liquid penetration into closed volumes and once more found a 
similar transition for the meniscus location as a function of the 
external driving forces. 

The rich behavior found above is certainly 
relevant to several microfluidic systems, where the
control of the liquid--gas or liquid--liquid interface at the small scale
is important. A generalization which fully takes into account the frequency
dependence of electric fields is a natural extension of the current work and 
should be explored \cite{tbj2,tbj3}, especially in the context of microfluidic systems.

 \section{Acknowledgments}
I am indebted to F. Brochard-Wyart and P. -G. de Gennes, with whom I had numerous fruitful
discussions
and correspondences on the subject.
For stimulating comments, I would like to thank A. Marmur and T. B. Jones.
This research was supported by the Israel Science Foundation (ISF)
Grant No. 284/05.




\providecommand{\refin}[1]{\\ \textbf{Referenced in:} #1}

\end{document}